\documentclass[aps,twocolumn, amsmath,amssymb, floatfix, citeautoscript]{revtex4}
\usepackage{graphicx, subfigure}
\usepackage{hyperref}
\usepackage{dcolumn}
\usepackage{bm}
\usepackage{color}

\newcommand{\comment}[1]{}

\newcommand{\Rs} {R_{s}}

\newcommand{\faco}{\left(1-\frac{1}{\epsilon}\right)}

\newcommand{\asymgbinv}{ \left(1 + sgn\left[\sum_j q_j e^{-\tau\frac{r_{ij}^2}{R_iR_j}}\right] \frac{\delta}{R_i-\Rs + \rho_w}\right)}

\begin{document}
\title{Accurate Evaluation of Charge Asymmetry in Aqueous Solvation}
\author{Abhishek Mukhopadhyay}
\affiliation{Department of Physics, Virginia Tech, Blacksburg, VA 24061, USA}
\author{Igor S Tolokh}
\affiliation{Department of Computer Science, Virginia Tech, Blacksburg, VA 24061, USA}
\author{Alexey V Onufriev}
\email{alexey@cs.vt.edu}
\affiliation{Department of Computer Science, Virginia Tech, Blacksburg, VA 24061, USA}
\affiliation{Department of Physics, Virginia Tech, Blacksburg, VA 24061, USA}
\email{alexey@cs.vt.edu}
\begin{abstract}
Charge hydration asymmetry (CHA) -- a characteristic dependence of
hydration free energy on the sign of the solute charge -- quantifies the
asymmetric response of water to electric field at microscopic level.
Accurate estimates of CHA are critical for understanding of
hydration effects ubiquitous in chemistry and biology. However, measuring
hydration energies of charged species is fraught with significant 
difficulties, which lead to unacceptably large (up to 300 \%) variation in
the available estimates of the CHA effect. We circumvent these difficulties
by developing a framework which allows us to extract and accurately
estimate the intrinsic propensity of water to exhibit CHA from accurate
experimental hydration free energies of neutral polar molecules.
Specifically, from a set of 504 small molecules we identify two pairs that
are analogous, with respect to CHA, to the K$^+$ /F$^-$ pair -- a classical
probe for the effect. We use these ``CHA-conjugate'' molecule pairs to
quantify the intrinsic charge-asymmetric response of water to the
microscopic charge perturbations: the asymmetry of the response is strong,
$\sim$ 50\% of the average hydration free energy of these molecules. The
ability of widely used classical 
water models to predict hydration energies of small
molecules correlates with their ability to predict CHA.
\end{abstract}
\maketitle
\section{Introduction}
The water molecule, H$_2$O, is among the simplest chemical structures, yet
the liquid phase of water is among the most complex liquids. Many of its
unique, vital for Life properties, including the ability to establish
complex hydrogen bonded structure~\cite{Marechal2006}, are due to the
complexity of electrostatic interactions. The ability of water to hydrate
ions and biomolecules is crucial to biological functions; detailed
atomistic understanding of these functions is impossible without accurate
description of the energetics of aqueous solvation (hydration). Given the
very polar nature of water as a solvent, understanding key details of
molecular hydration requires accurate and comprehensive experimental
characterization of electrostatic properties of water molecule in liquid
phase at microscopic level. This characterization is currently lacking: for
example, even the value of the dipole moment of water molecule in
liquid phase is still debated\cite{Leontyev2012}: the experimental range is from
2.7 to 3.2D,~\cite{Silvestrelli1999,Sharma2007,Badyal2000} and higher
multipole moments are not available from experiment at all. 

\begin{figure}[htbp]
\centering
\includegraphics[trim=0cm 6.5cm 1cm 2.5cm, clip=true, scale=0.32]{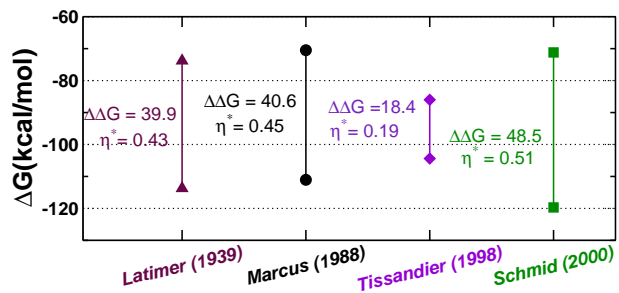}
\caption{Uncertainty in experimentally determined charge hydration asymmetry
(CHA) for K$^+$/F$^-$ ion pair. Shown are the hydration free energies, 
$\Delta G$, for
K$^+$(top markers) and F$^-$(bottom markers) from four classical experimental
references~\cite{Latimer1939,Marcus1988,Tissandier1998,Schmid2000}, 
from historical to modern, spanning more than eight decades. Also shown are
the absolute CHA, $\Delta \Delta G=\Delta G(\mbox{K}^+) - \Delta G(\mbox{F}^-)$, and
the dimensionless relative CHA~\cite{Mukhopadhyay2012}, which quantifies the 
asymmetry of the charge hydration:
$\eta^*=\left|\frac{\Delta \Delta G}{\left<\Delta G\right>}\right|$, where 
$\left<\Delta G\right>=\frac{1}{2}\left(\Delta G(\mbox{K}^+)+\Delta G(\mbox{F}^-)\right)$.}
\label{fig:KF}
\end{figure}

Experiments further lack consensus in quantifying the key characteristic of
the asymmetry of water response to microscopic fields of solvated charges
-- the so called charge hydration asymmetry
(CHA)~\cite{Buckingham1957,Rashin1985,Hirata1988,Roux1990,Hummer1996,Lynden_bell1997,Rajamani2004,Grossfield2005,Mobley2008,Purisima2009,Mukhopadhyay2012,Bardhan2012,Mukhopadhyay2014,Scheu2014,Bardhan2014},
Fig.~\ref{fig:KF}. One of its earliest~\cite{Latimer1939} known manifestations
is the observed differences in hydration free energies of oppositely
charged ions of similar size, {\it e.g.}, the K$^+$/F$^-$
pair, Fig.~\ref{fig:KF}; completely unaccounted for in the linear response
continuum framework~\cite{Born1920}. These CHA-related differences are larger than many
relevant biomolecular energy scales, such as folding free energy of a
typical protein. The CHA effects are strong not only for the charged, but
also for net neutral solvated
structures~\cite{Mobley2008,Mukhopadhyay2014,Bardhan2014}.
Absolute single-ion solvation free energies are conventionally deduced
from experimental estimates of hydration enthalpies of ionic solutions
via the Born-Haber cycle~\cite{Tissandier1998,Schmid2000} using highly
controversial estimation of absolute proton (H$^+$) hydration enthalpy.
Uncertainties of up to $16$ kcal/mol are reported that stem from various
degrees~\cite{Donald2011} of extra-thermodynamic
assumptions~\cite{Marcus1988,Schmid2000,Hunenberger2011} or cluster ion solvation
approximation~\cite{Tissandier1998} made in obtaining these estimates. 
Furthermore, ion hydration free energy is directly affected by 
liquid-vapor and
liquid-cavity surface (interface) potentials: every time a charge $q$ crosses 
such interface, the electrostatic energy of the charge 
changes by $q \phi_{int}$, where $\phi_{int}$ 
is the electrostatic potential difference associated with a the interface. 
However, available experimental estimates of this
elusive quantity can differ by a factor of five~\cite{Farrell1982,Fawcett2008};
there are fundamental difficulties associated with
measuring and even interpreting~\cite{Kathmann2011} the water surface
potential~\cite{Pethica2007,You2014}.

	The net result of these, and possibly other problems,  
is an almost 300\% variation between the four available comprehensive
experimental data sets of ion hydration energies in the strength of the
CHA effect for K$^+$/F$^-$ pair, Fig.~\ref{fig:KF}.
At the same time, even uncertainties of
the order of just several $k_BT$ can have profound effect on the
thermodynamics or kinetics of different phenomena where ion dehydration
is involved, e.g., membrane ion channel transport, ion binding to
receptor proteins, etc. An accurate quantification of the inherent CHA
of water is therefore of paramount importance for accurate quantitative
estimates in biology, chemistry, and physics.

In contrast to ions, direct measurements of experimental hydration
energies~\cite{Rizzo2006,Mobley2009} of neutral solutes are relatively
straightforward and very accurate, with errors within a fraction of
kcal/mol.  The two primary sources of error in the case of ionic solutes
are completely eliminated in the case of net neutral solutes, $\sum_i
q_i=0$. First, the contribution of the interface  potential to the energy
of the solute transfer  is zero, $\phi_{int}\sum_i q_i = 0$.  Second,
experimental hydration energy estimates  of  neutral molecules are
typically obtained using calorimetric measurements of transfer free energy,
which are free from empirical reference energies such as the proton
hydration  free energy needed in the case of ions.~\cite{Abraham1990}
Connection to computational models is also easy to make since the
relationship between measured  and computed hydration energies is
straightforward for neutral
solutes~\cite{Mobley2009,Roux1999,Mobley2012,Roux2014,Dill2014}, in
contrast to ions, see {\it e.g.} Ref.~\cite{Roux2014} for a detailed
overview of the associated  issues.  We therefore propose to quantify the
CHA effect by utilizing experimental hydration energies of small neutral
molecules, rather than ions. Specifically, we identify ``CHA-conjugate"
pairs of anion-like and cation-like neutral polar molecules that behave
just like the K$^+$/F$^-$ pair with respect to charge hydration asymmetry.
To characterize the asymmetry of water response to microscopic charge
independently of the strength of the charge probe (absolute hydration
energies of small neutral molecules are much smaller than those of ions) in
what follows we will quantify the strength of the CHA effect by a
dimensionless quantity $\eta^*=\left|\frac{\Delta\Delta G}{\langle \Delta
G\rangle}\right|$, where  $\langle \Delta G \rangle$ is the average
hydration energy of the pair of solutes, Fig.~\ref{fig:KF}.

\section{Results and Discussion}
Without loss of generality, the hydration free energy, $\Delta G$, of
a molecule can be decomposed into symmetric, $\Delta G^{sym}$, and
asymmetric, $\Delta G^{asym}$, parts with respect to inversion of solute charges:
$\Delta G = \Delta G^{sym} + \Delta G^{asym}$; such that upon sign
inversion of the partial atomic charges $\Delta G^{sym}$ remains
invariant, whereas $\Delta G^{asym}\rightarrow -\Delta G^{asym}$.
Physically, the proposed symmetric-asymmetric decomposition
of the solvation free energy reflects the inherently charge-asymmetric
microscopic response of water to perturbing electric
field due to a solvated charge. Here, $\Delta G^{asym}$ part accumulates
all the deviations from the symmetric response $\Delta G^{sym}$
that is invariant upon inversion of the perturbing electric field;
the response due to dipole polarization of water contributes primarily 
to the polar part of $\Delta G^{sym}$. The non-polar component of $\Delta G$
for molecules is also mostly symmetric~\cite{Mobley2008}, and is similar 
in magnitude for molecules of similar size.
The magnitude of $\Delta G^{asym}$ relative to $\Delta G^{sym}$
characterizes charge asymmetry of the solvent response: {\it e.g.} 
in a hypothetical water consisting of molecules with a perfectly tetrahedral
charge distribution this asymmetric response is zero\cite{Mukhopadhyay2012}.

For an arbitrary pair (A,B) of molecules, the experimental difference
in their hydration energies $\Delta \Delta G = \Delta G(A) - \Delta G(B)$ 
would contain a mix of CHA-related and unrelated components. Here
we propose to cleanly isolate the asymmetric part by identifying special
pairs of ``CHA-conjugate'' neutral molecules -- molecules with same $\Delta
G^{sym}$, but equal and opposite $\Delta G^{asym}$; for these special
pairs $\Delta G^{sym}$ would cancel out in the difference, while $|\Delta
G^{asym}|$ would combine, so that $\Delta \Delta G=\Delta G^{asym}(A) - \Delta G^{asym}(B)$ would contain only
the CHA effect we seek to quantify.

Since no known experimental method can perform the decomposition of 
the total solvation energy into charge-symmetric and asymmetric parts, 
we must resort to theoretical water models to identify
such special pairs of molecules. 
However, once such a pair is identified, the values of $\Delta G^{sym}$ 
and $\Delta G^{asym}$ for each molecule in the A/B pair
can be extracted from the experimental hydration energies;
$\Delta G^{sym} = \frac{1}{2}(\Delta G(A) + \Delta G(B))$ and 
$|\Delta G^{asym}| = \frac{1}{2}|\Delta G(A) - \Delta G(B)|$.
The implications of the decomposition for development of simplified 
water models will be discussed below. 
\begin{figure*}[htbp]
\centering
\includegraphics[scale=0.18]{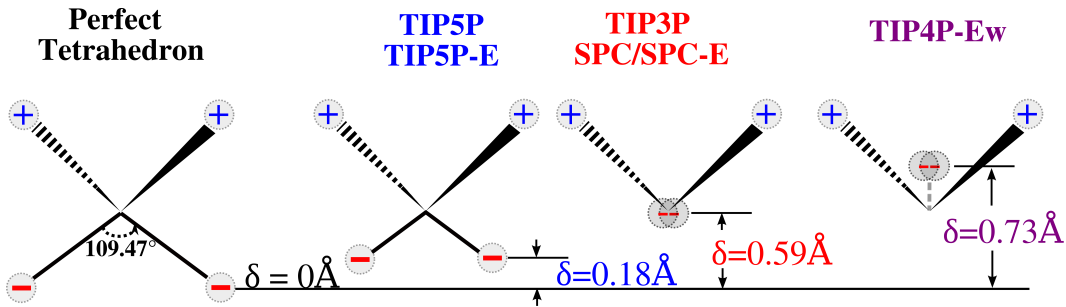}
\caption{A family of rigid $n$-point water models arranged in ascending
order with respect to deviation from the perfect tetrahedral
symmetry (BNS model, zero CHA). The ability of the water models to
cause CHA is proportional to the magnitude of parameter $\delta$
that progressively breaks the specific charge inversion symmetry in
this family of models.}
\label{fig:watmod}
\end{figure*}

Within the widely used family of TIPnP water models~\cite{Mahoney00},
the propensity to cause CHA increases~\cite{Rajamani2004,Mobley2008}
progressively from TIP5P to TIP3P\cite{TIP3P} to TIP4P-Ew\cite{TIP4PEW}.
In contrast, one of the earliest water models, BNS\cite{Rahman1971}, 
does not exhibit any CHA: due to its perfectly tetrahedral geometry, 
charge inversion within this molecule can be emulated by a set of rotations
around its oxygen center -- a sufficient 
condition for the absence of CHA~\cite{Mukhopadhyay2012}. This high
degree of charge inversion symmetry is progressively broken in other water models 
shown in Fig.~\ref{fig:watmod} by an order parameter $\delta$ that has an
intuitive geometric interpretation.
A relation between $\delta$ (geometry) and CHA (energy) is established
in SI, where it is shown that $\delta
\sim \Omega_2$ -- the cubic octupole moment of water molecule 
earlier recognized~\cite{Kusalik1988b,Ichiye2010,Ichiye2011,Mukhopadhyay2012}
as key for breaking the charge hydration symmetry.
This multipole moment is defined as 
$\Omega_2 =(1/2)(\Omega_{yyz}-\Omega_{xxz})$ where $\Omega_{ijk}$
are the components of the traceless octupole tensor of a water molecule model
in the Cartesian frame with origin at the water oxygen center
and OH bonds in the $yz$ plane with $z$-axis bisecting the H--O--H 
bond angle.

\begin{figure}[htbp]
\centering
\includegraphics[trim=0cm 0cm 0cm 8cm, clip=true, scale=0.32]{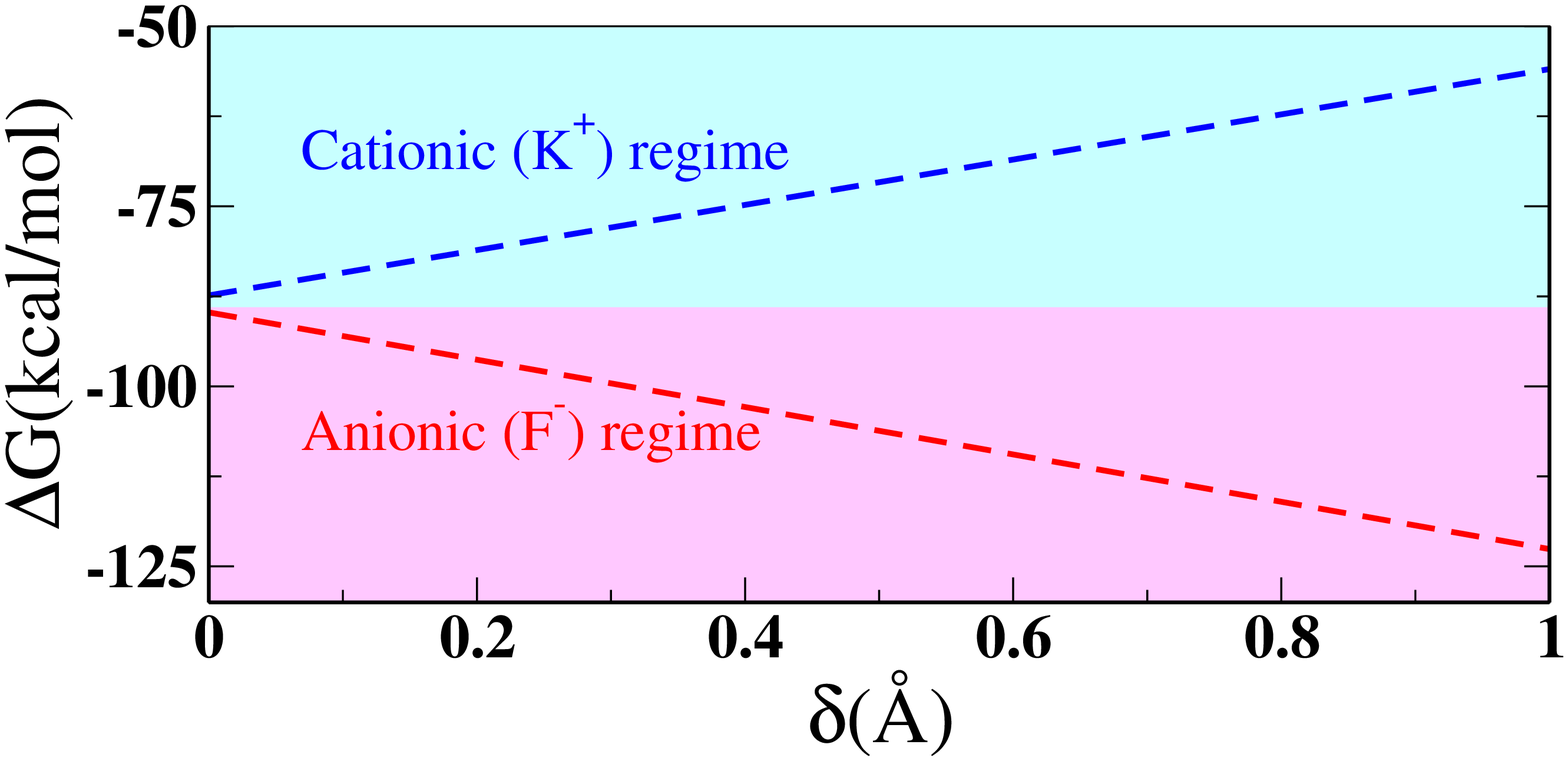}
\caption{K$^+$ (blue dashed line) and F$^-$ (red dashed line) ion pair
defines the characteristic behavior of CHA-conjugate pair of solutes -- a symmetric,
monotonically increasing difference in the electrostatic hydration free energies
with respect to increasing degree of the perturbation $\delta$ that
breaks the charge-inversion symmetry of the water models. Hydration energies are
computed via ``CHA-aware'' Born equation\cite{Mukhopadhyay2012}, 
\ref{eq:bornreborn}, with $R_s = 0.52$ \AA, $R_w = 1.4$~\AA, 
and ionic radii from Ref.\cite{Marcus1991} }
\label{fig:KFdg}
\end{figure}

The asymmetric part of the solvation free energy,
$\Delta G^{asym}$ can be treated as a first order perturbation (with
respect to the symmetry breaking variable $\delta$) to the completely
charge-symmetric part, $\Delta G^{sym}=  \Delta G (\delta = 0)$; {\it i.e.}
$\Delta G(\delta) \simeq \Delta G (\delta = 0) + \frac{\partial \Delta
G}{\partial \delta} \delta$. For small $\delta$, $\Delta G^{asym}\simeq 
\frac{\partial \Delta G}{\partial \delta} \delta \propto \delta$. To
illustrate the idea within a concrete
theoretical framework we follow an earlier work~\cite{Mukhopadhyay2012}
where CHA is introduced explicitly into the Born equation~\cite{Born1920}:

\begin{equation}
\Delta G \simeq -\left(1-\frac{1}{\epsilon}\right)\frac{q^{2}}{2(R +
R_s)} \left(1 - sgn[q] \frac{\delta}{R + R_w}\right) \, .
\label{eq:bornreborn}
\end{equation}
Here, $\epsilon$ is the dielectric constant of water, $R_w$ is the radius
of water molecule, $q$ and $R$ are the ion charge and ionic radius,
respectively, and $R_s$ is a constant shift to the dielectric 
boundary~\cite{Mukhopadhyay2012}.
Neglecting a minor difference in the ionic radii
of K$^+$ and F$^-$ ions, \ref{eq:bornreborn} results in
the same $\Delta G^{sym}=\Delta G$($\delta =0$) for both ions and
a symmetric gap 
in hydration free energies 
\begin{equation}
\Delta \Delta G(\delta) = 
2|\Delta G^{asym} (\delta)| = 
\left(1-\frac{1}{\epsilon}\right)\frac{q^{2}}{(R + R_s)} \frac{\delta}{(R + R_w)} \, , 
\label{eq:gap}
\end{equation}
which  increases monotonically with $\delta$, Fig.~\ref{fig:KFdg}.
In other words, for a cation/anion pair of the same size,
\ref{eq:bornreborn} describes a ``CHA-conjugate'' ion pair with identical
$\Delta G^{sym}$ and equal but opposite $\Delta G^{asym}$.

We propose to use these
two distinct characteristics of the $\Delta G (\delta) $ for a pair of 
a ``CHA-conjugate'' solutes
as a signature feature to identify CHA-conjugate pairs
among neutral molecules. That is, we seek pairs of
small polar neutral molecules for which $\Delta G(\delta)$ behaves similar
to that of the K$^+$/F$^-$ pair in Fig.~\ref{fig:KFdg}, despite differences in scales
of hydration free energies -- hydration free energies of small neutral molecules are 
an order of magnitude less than that of the ionic pair. 
In hydrated polar molecules, for oppositely partially charged atoms  
one can expect local cation-like or anion-like response of water.
The contributions to the overall CHA from the hydration of these           
atoms, especially atoms of the polar groups of a molecule,
may be quite dissimilar. In general, these contributions
will not cancel each other completely, resulting in a net CHA effect.
In addition to that, in some pairs of molecules, these contributions
may combine in a way to produce the $\Delta G(\delta)$ pattern we seek,
Fig.~\ref{fig:KFdg}.   
Note that the very existence of such pairs is not guaranteed
{\it a priori}: hydration energy of a molecule, as well as its symmetric
and asymmetric parts,
are influenced by a complex interplay of screened interactions between its
partial charges. To illustrate the above points mathematically  
we use the multi-atomic analog of \ref{eq:bornreborn}, derived directly
from the so-called CHA-GB model~\cite{Mukhopadhyay2014},~\ref{eq:chagb} in
Computational Methods.
Within this formalism, 
\begin{equation}
  \Delta G(\delta)  \simeq  
  -\frac{1}{2}\faco\sum_{i,j}\frac{q_iq_j}{f_{ij}^{GB}}
  \left[1-\delta \mathcal{F}_{ij} \right ] \, , 
 \label{eq:dgcha}
\end{equation}
where $\frac{1}{f_{ij}^{GB}} =
\left(r_{ij}^2+R_iR_je^{-\frac{r_{ij}^2}{4R_i R_j}}\right)^{-1/2}$
is the purely charge-symmetric Green function of the generalized Born
formula~\cite{Still1990,Onufriev2010}, and 
$\mathcal{F}_{ij} = \left[ \frac{R_i R_je^{-\frac{r_{ij}^2}{4R_i R_j}}}{{f_{ij}^{GB}}^2}\left(\frac{sgn[\sum_kq_k e^{-\tau \frac{r_{ik}^2}{R_i R_k}}]}{R_i-R_s+R_w} + \frac{sgn[\sum_lq_l e^{-\tau \frac{r_{jl}^2}{R_j R_l}}]}{R_j-R_s+R_w}\right)   \right]  $. 
Note that \ref{eq:dgcha} has the same structure as \ref{eq:bornreborn}
with respect to variation in $\delta$, which further rationalizes the
use of our ion-inspired criteria, Fig.~\ref{fig:KFdg}, for identifying 
CHA-conjugate pairs among molecules. However, in contrast to the case of a pair
of oppositely charged monovalent ions of the same size, for most pairs 
of neutral molecules their $|\Delta G^{asym}|$ are not equal to each other even
if their $\Delta G^{sym}$ are the same. This is due to a complex interplay
between the self ($\mathcal{F}_{ii}$) and the cross (charge-charge
interactions, $\mathcal{F}_{ij}$) contributions to CHA within each
molecules. In fact, as we show later, we found only two pairs of
CHA-conjugate molecules from an available comprehensive diverse 
set of $504$ small,
neutral molecules~\cite{Mobley2009} with known experimental hydration
free energies (see Computational Methods). Note that these experiments
were conducted under conditions to ensure~\cite{Rizzo2006} that these
molecular solutes were indeed neutral, and therefore avoid the issues
related to uncertainties in hydration energies of charged solutes
discussed earlier. The fast analytical CHA-GB model~\cite{Mukhopadhyay2014}
was used in the initial screening for the signature
$\Delta \Delta G (\delta)$ gap, and the best candidates were then confirmed
by careful free energy perturbation (FEP) calculations~\cite{Mobley2009}
using the three TIPnP explicit water models in Fig.~\ref{fig:watmod}, see
Computational Methods.

	In the search for CHA-conjugate pairs of small molecules, 
we utilize rigid, non-polarizable
water models of the same geometry (TIPnP family) that have roughly the same 
dipole moment (less than 0.06 D difference between the models) 
and exhibit a well studied
behavior with respect to CHA \cite{Mobley2008,Mukhopadhyay2012}. 
It was shown that the different CHA effects these models can
produce are proportional to the values of their cubic octupole moments
$\Omega_2$ and thus to the different values of 
the symmetry breaking parameter $\delta$ inherent to these models.
It is also critical that the monotonic increase of the $\Delta \Delta G(\delta)$ gap 
with $\delta$ is symmetrical, that is
$\Delta G^{sym}$ is practically the same within this family of water models.
These two properties of TIPnP family are essential for the identification of
CHA-conjugate pairs of solutes.

\begin{figure}[htbp]
\centering
\includegraphics[trim=0cm 0cm 1cm 5cm, clip=true, scale=0.32] {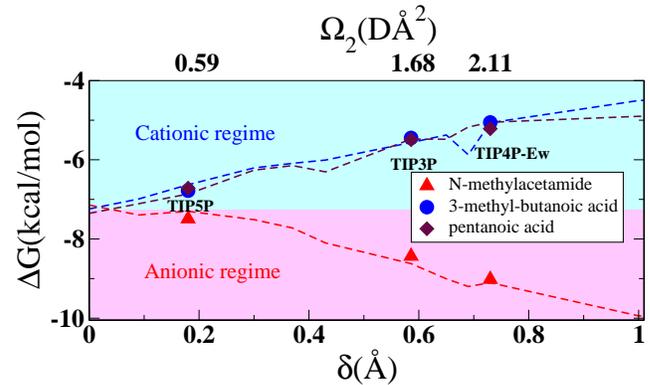}
\caption{ Hydration free energies of two pairs of 
molecules analogous to K$^+$/F$^{-}$ pair with respect to CHA; 
compare with Fig.~\ref{fig:KFdg}. The energies are computed using the CHA-GB model
(dashed lines)~\cite{Mukhopadhyay2014}, and free energy perturbation (FEP) calculations
(solid markers), see Computational Methods. Statistical error bars (not shown) for
the FEP calculations are smaller than the symbol size}
\label{fig:goodpairs}
\end{figure}

\begin{figure}[htbp]
\centering
\includegraphics[scale=0.25]{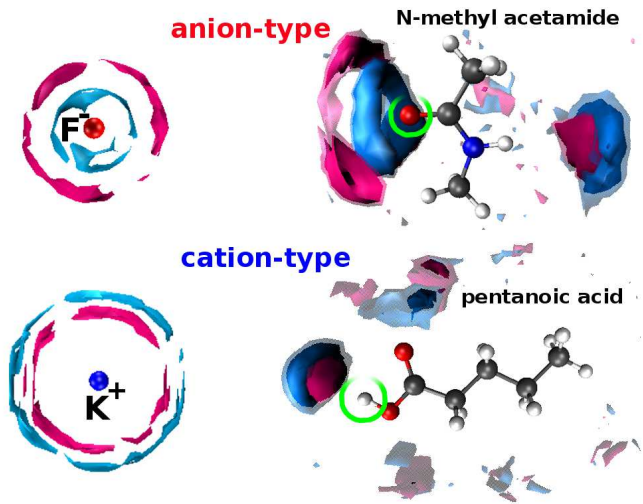}
\caption{ Similarity between water-structuring effects of ions (left)
and one pair of CHA-conjugate neutral polar molecules (right). Shown are
water-oxygen (magenta) and water-hydrogen (cyan) occupancy maps computed
using TIP3P water model. High probability isomaps are shown using
solid surfaces and relatively lower probability isomaps are shown
as translucent surfaces. Atom that contributes predominantly to the
CHA effect in each molecule is pointed out by green circle. 
The other cation-like molecule, 3-methyl
butanoic acid, is an isomer of pentanoic acid and has similar occupancy
isomap, see SI. Using TIP4P-Ew
instead of TIP3P results in similar occupancy maps (not shown). } 
\label{fig:efld}
\end{figure}

	Our search has yielded two CHA-conjugate pairs of neutral, but highly
polar molecules, Fig.~\ref{fig:goodpairs}, which by our quantitative 
criteria behave just like the K$^+$/F$^-$ ion pair with respect to CHA, 
compare Fig.~\ref{fig:goodpairs} and Fig.~\ref{fig:KFdg}. 
The qualitative rationalization for the unique behavior
of these two pairs with respect to CHA is illustrated for one of them in
Fig.~\ref{fig:efld}. 
The anion-like (cation-like) molecule
has one relatively highly charged, ``CHA-dominant'' atom, {\it e.g.} the
amide oxygen of N-methyl acetamide (carboxylic hydrogen of pentanoic
acid) that behaves like F$^-$ (or K$^+$) with respect to the structuring
of water around it. The effect of the opposite charges within each
molecule is much more diffuse, and can not cancel the dominant
asymmetric contribution from this one atom.

We quantify the charge hydration asymmetry of these two pairs by their 
{\it experimental} relative CHA,
$\eta^*=\left|\frac{\Delta \Delta G}{\langle\Delta G\rangle}\right|$.
The experimental $\eta^*$ for the two CHA-conjugate pairs of neutral small 
molecules (3-methyl butanoic acid, N-methylacetamide) and 
(pentanoic acid, N-methylacetamide), 
respectively, are presented in Table~\ref{tab:eta0}.
%
\begin{table}[htbp]
   \caption{ Relative CHA, $\eta^*$, of two CHA-conjugate pairs of
neutral molecules from the experimental hydration free energies~\cite{Rizzo2006,Mobley2009}.}
\resizebox{0.5\textwidth}{!}{%
  \begin{tabular}{@{\vrule height 10.5pt depth4pt
    width0pt}c|c}
    \hline
    Molecule Pair&Experimental relative CHA, $\eta^*$\\
    \hline
\includegraphics[scale=0.08]{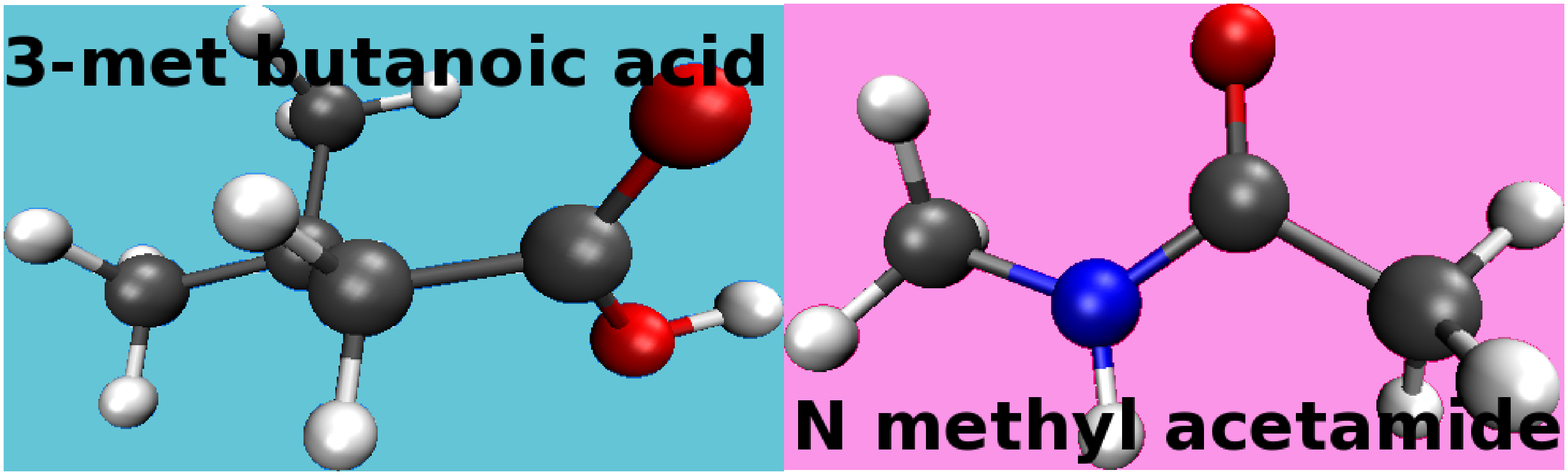}&0.49\\
\hline
\includegraphics[scale=0.08]{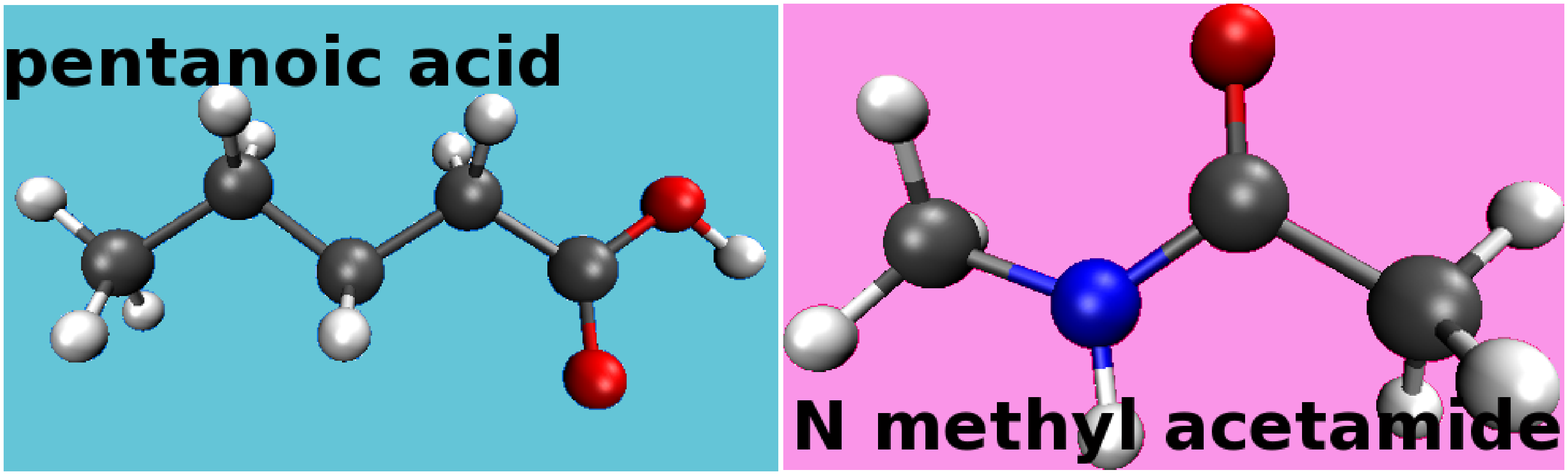}&0.48\\
  \hline
\end{tabular}}
   \label{tab:eta0}
\end{table}

The physical meaning of the large, almost $0.5$, experimental value of
$\eta^*$ is that the intrinsic charge-asymmetric response of water to
microscopic charges (atomic size CHA perturbations) is strong,
comparable to the charge hydration energy itself. For example, 
experimental hydration energies of 
pentanoic acid and N-methylacetamide
are -6.16 and -10.00 kcal/mol respectively,~\cite{Rizzo2006} {\it i.e.} the
$|\langle\Delta G\rangle|=$ 8.08 kcal/mol while the
$\Delta\Delta G$= 3.84 kcal/mol for the pair.
In fact, the asymmetry
of the individual perturbation caused by a single ``CHA-dominant'' atom 
is even stronger, as the net CHA effect in a pair of molecules measured
by $\eta^*$ can be attenuated somewhat by the opposing contributions from
other charges, see Fig.~\ref{fig:efld} and SI. We again stress that the
$\eta^*$ from Table~\ref{tab:eta0} are free of the issues related to
the fundamental or technical difficulties~\cite{Roux2014} associated
with the ion hydration asymmetry: $\eta^*$ inferred from CHA-conjugate pairs
of neutral polar molecules directly and accurately characterize the
intrinsic asymmetry of the water response to microscopic electric field.
As expected, the uncertainty in the experimental $\eta^*$ is small,
about 2 \%, as seen from the difference between the two
CHA-conjugate pairs. This high accuracy 
is in contrast to the almost 300 \% difference between
available experimental hydration energies for K$^+$/F$^-$ pair, Fig.~\ref{fig:KF}.

\begin{figure}[htbp]
\centering
   \includegraphics[width=5.5cm]{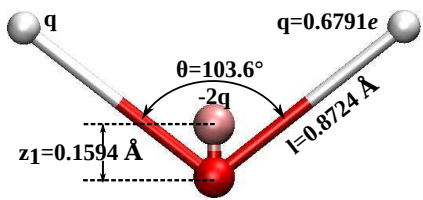}
   \caption{ Charge distribution in OPC water model\cite{Anandakrishnan2013,Izadi2014}.
The distance between the oxygen center and the positive charges is noticeably shorter 
than the $\sim$ 1 \AA~ assumed by common 3 point
charge water models\cite{TIP3P,TIP4PEW}.}
   \label{fig:OPC}
\end{figure}

	Looking through the prism of CHA at the development of classical
fixed-charge water models during the past several decades, we note that,
geometrically, the largest variation between the popular water models
appears along the single coordinate: one that breaks the perfect
charge-inversion symmetry of a perfect tetrahedral charge arrangement
assumed by early BNS model, Fig.~\ref{fig:watmod}.
Judging by the ability of each water model in Table~\ref{tab:eta} to
predict the experimental $\eta^*$, we conclude that to the extent that
the charge distribution of a real water molecule in liquid water can be
approximated by three point charges, the distribution might be close to the recently
proposed water model OPC~\cite{Anandakrishnan2013,Izadi2014}, Fig.~\ref{fig:OPC}. 
%
\begin{table}[htbp]
   \caption{ Relative CHA, $\eta^*$, of two CHA-conjugate pairs of
neutral molecules from experiment and
explicit water simulations. Water models used for comparison are TIP5P, TIP3P,
TIP4P-Ew and a recent 4-point model OPC~\cite{Anandakrishnan2013,Izadi2014}}
\resizebox{0.5\textwidth}{!}{%
  \begin{tabular}{@{\vrule height 10.5pt depth4pt
    width0pt}c|c|c|c|c|c}
    \hline
    Molecule Pair&Exp&\multicolumn4c{Water models}\\
    \cline{3-6}
    &&TIP5P&TIP3P&TIP4P-Ew&OPC\\
    \hline
\includegraphics[scale=0.08]{./figures/pair1.eps}&0.49&0.10&0.43&0.56&0.45\\
\hline
\includegraphics[scale=0.08]{./figures/pair2.eps}&0.48&0.11&0.42&0.53&0.45\\
  \hline
\end{tabular}}
   \label{tab:eta}
\end{table}
%
This new water model is, perhaps, the very first attempt to explicitly
consider the CHA effect in water model optimization~\cite{Izadi2014};
the model is curiously successful in accurately reproducing most
properties of liquid water~\cite{Izadi2014}. Importantly, the accuracy
of the predicted small molecule hydration energies by the 
water models used in this study (in Table \ref{tab:eta}) correlates well with their ability to reproduce
experimental CHA. Namely, RMS errors against experiment in hydration
free energy estimates for TIP5P, TIP3P, TIP4P-Ew and OPC are,
respectively: 1.85, 1.10, 1.14, and 0.97 kcal/mol (based on 20 neutral molecules, 
see Computational Methods, chosen~\cite{Izadi2014} to span the entire hydration energy
range of the set of 504 neutral molecules). The correlation suggests that the
highly accurate experimental $\eta^*$ for the CHA-conjugate pairs of
neutral small molecules can be used to calibrate and improve water
models, especially their ability to accurately describe microscopic
electrostatic effects important for molecular hydration. 
The relative quantity $\eta^*$ is expected to be insensitive to parameters
describing the molecules in ``CHA-conjugate'' pair (such as the specific
parametrization of atomic partial charges) meaning that predicted $\eta^*$ 
mostly characterizes propensity of the water model to cause CHA.  
This propensity is strong, 
consistent with the large value of the
charge-inversion symmetry breaking separation $\delta$,
Fig.~\ref{fig:watmod}, comparable to the positive-negative charge
separation in water models that approximate the experimental CHA
reasonably well, Table~\ref{tab:eta}. The CHA strength is also
consistent with a relatively large value of the $\Omega_2$ octupole
moment of water molecule in liquid phase predicted by several
independent quantum mechanical estimates\cite{QM230TIP5P,Ichiye2011}.

As we have noted before, for each CHA-conjugate molecule pair 
one can extract $\Delta G^{sym}$ directly from the experimental hydration free energies,
which is essentially the
average hydration energy of the pair. This quantity can also be used to assess
the quality of water models with respect to their ability to describe the symmetric
response to solvated charge. For example, for the pairs in Table~\ref{tab:eta},
OPC water model gives the best agreement $\Delta G^{sym}\simeq$ -7.3
kcal/mol for each pairs, vs. the experimental value of $\simeq$ -8
kcal/mol; $\Delta G^{sym}$ for other
water models tested here are less accurate, Fig.~\ref{fig:goodpairs}. In the
future, it would be interesting to test other water models, including 
polarizable ones, with respect to their ability to reproduce experimental
$\eta^*$ and $\Delta G^{sym}$. We expect that for the current
polarizable models, for which the polarizability is added at the dipole
level~\cite{Lamoureux2003,Gunsteren2005,Wang2012,Leontyev2012},  
one would observe similar $\eta^*$ to the corresponding non-polarizable ``base''
model to which the polarizability is added. This is because the dipole
moment response is symmetric with respect to solute charge inversion.

Having quantified the intrinsic CHA effect of water by $\eta^*$ for a
pair of neutral, cation-like and anion-like small molecules, we inquire
if we can use the accurate experimental value of $\eta^*$ 
to set a lower bound on the value of ${\eta^*}^{ion}$ for the
(K$^+$/F$^-$) pair and thus reduce the dramatic uncertainty associated
with the original, ionic CHA shown in Fig.~\ref{fig:KF}. 
To this end, we decompose the molecular $\eta^*$ onto individual atomic
contributions. To set the bound on $\eta^*$ we single out the
``CHA-dominant atom'' for each molecule, Fig.~\ref{fig:efld}, and use the
fact that the CHA effect of these two atoms is larger than that of
the combined effect from all the atoms in the molecular pair.
This is because of partial cancellation of CHA for neutral molecules.
Note that if the CHA-dominant atoms for these CHA-conjugate molecule
pairs were exactly of the same size as those of the K$^+$/F$^-$ pair, then
$\eta^*$ from Table~\ref{tab:eta0} could be used directly to set the
lower bound on the $\eta^{*ion}$. In reality, the radii of the CHA-dominant
``cation'' or ``anion'' atoms, Fig.~\ref{fig:efld}, are 
different from that of K$^+$ or 
F$^-$, see SI. Nonetheless $\Delta \Delta G$ scales with the ion
size 
according to~\ref{eq:gap}, which is consistent with
experiment~\cite{Marcus1988,Schmid2000,Tissandier1998}. We use this 
simple scaling relationship to relate atomic (CHA-dominant atoms of
the molecules) and ionic $\eta^*$, see SI.    
	To obtain more refined estimate of the bound,
%
one needs to consider additional contributions specific to ionic
$\Delta \Delta G$, specifically the non-polar 
interactions estimated in e.g. Ref. ~\cite{Duignan2013} and
water-vapor interface potential. 
For the oppositely charged ions, both the liquid-vapor interface
potential~\cite{Kathmann2011} and the ion cavity potential~\cite{Ashbaugh2000}
contribute to $\Delta \Delta G$: the net contribution is $|2e|\phi_{int}$,
where $\phi_{int}$ is the (positive) electrochemical potential corresponding to
taking an ion from vacuum to bulk water. Note that the contribution
is positive for the positive potential, and thus can only increase
${\eta^*}^{ion}$ relative to molecular $\eta^*$. Experimental 
estimates of $\phi_{int}$ range from +25 mV~\cite{Farrell1982} 
to +140 mV~\cite{Fawcett2008}.  For the
lower bound estimate of ${\eta^*}^{ion}$, taking the lower bound on the
experimental estimate of $\phi_{int} = 25$ mV, 
results in a negligible contribution to $\Delta \Delta G$.\\
	The end result is a lower bound of ${\eta^*}^{ion}(\mbox{K}^+/\mbox{F}^-)$ 
set by the molecular $\eta^*$:
${\eta^*}^{ion}  > 0.50\eta^* +0.14$. Here, $0.14$ is the non-polar 
contribution to ${\eta^*}^{ion}$, 
which, according to a recent theoretical estimate\cite{Duignan2013} 
we use here,  
is not negligible for the K$^+$/F$^-$ ion pair, and originates mostly 
from the difference in ion-water dispersion interactions
for these ions.
Substituting  ${\eta^*} = 0.48$ from Table~\ref{tab:eta0}, we arrive at:  
\begin{equation}
{\eta^*}^{ion} (\mbox{K}^+/\mbox{F}^-) >0.38
\label{eq:compare2}
\end{equation}
This lower bound estimate suggests that 
the ``original'' relative CHA effect for the K$^+$/F$^-$ ion
pair, Fig.~\ref{fig:KF}, is also likely to be strong. 
	
To conclude this section, we propose an experiment that can, in 
principle, estimate $\Delta \Delta G$ and $\eta^*$ of K$^+$/F$^-$ pair 
accurately, avoiding the uncertainties discussed earlier, Fig.~\ref{fig:KF}.  
The idea \cite{Kalidas2000} is to measure hydration energies of two net neutral 
ion pairs, K$^+$M$^-$ and L$^+$F$^-$, where L$^+$ and M$^-$ are relatively
large ions (compared to the K$^+$, F$^-$ ions) of similar size.
For such large ions, $|\Delta G(\mbox{L})| \sim |\Delta G(\mbox{M})| \ll |\Delta G(\mbox{K}^+)|, 
|\Delta G(\mbox{F}^-)|$. Thus, in very dilute aqueous solutions of K$^+$M$^-$
and L$^+$F$^-$, $\Delta G(\mbox{K}^+\mbox{M}^-) - \Delta G(\mbox{L}^+\mbox{F}^-) =
\Delta G(\mbox{K}^+) - \Delta G(\mbox{F}^-) + ( \Delta G(\mbox{M}^-) -
\Delta G(\mbox{L}^+) ) \simeq \Delta G(\mbox{K}^+) - \Delta G(\mbox{F}^-)=\Delta \Delta  G(\mbox{K}^+/\mbox{F}^-)$.
Once $\Delta \Delta  G(\mbox{K}^+/\mbox{F}^-)$ is quantified in 
this manner, accurate $\eta^* = \Delta \Delta  G(\mbox{K}^+/\mbox{F}^-)/\langle\Delta G\rangle$ 
can be obtained since the average $\langle\Delta G\rangle$ for the ion pair 
can be quantified very accurately. 

	Note that a variation on the above idea might be used to develop 
an alternative procedure for parameterization of ion force-fields for 
classical simulations\cite{Joung2008,Gladich2010,Li2015}. 
The first step would involve experimental determination of 
hydration energy of a salt
$n$F$^-$L$^{n+}$ where L$^{n+}$ is a large ``composite" ion such that 
for its surface atoms 
the force-field parameters are well established and are not related 
to small ion parameters, {\it e.g.}
Co(NH$_{3})_{6}^{+3}$ (cobalt hexammine) or Ni(NH$_{2})_{6}^{+2}$ ion. 
The next step would be to perform a series of explicit solvent 
simulations of $n$F$^-$ L$^{n+}$ neutral pair to fit F$^-$ parameters to 
match experimental hydration energy of  
$n$F$^-$L$^{n+}$ salt for each water model of interest. 
Once F$^-$ parameters are known, one can
perform a set of simulations to obtain cation ({\it e.g.} K$^+$)
parameters by fitting against known ion pair hydration energies. 
With the known cation
parameters, the procedure can be continued to estimate the parameters for other
anions. The oulined protocol would avoid a number of difficulties 
associated with both experimental 
and computational\cite{Roux2014} 
estimates of hydration energies of charged species. 

\section{Conclusions}
In this article we have proposed an approach to use highly accurate
experimental hydration energies of small neutral molecules to quantify
the effect of charge hydration asymmetry (CHA) -- the charge-asymmetric
response of water to microscopic electric field.  Characteristic
dependence of hydration free energies on the sign of charged solutes of
similar radii such as the K$^+$/F$^-$ pair is the best known
manifestation of CHA. However, for ion pairs, the quantification of the
hydration free energy, and hence the associated CHA, is highly uncertain
($\sim$ 300\% difference between four available comprehensive sets of experimental
data) due to a variety of fundamental, and technical
difficulties~\cite{Donald2011,Kathmann2011,Tissandier1998,Kelly2006,Roux2014}.
Here we overcome these difficulties and accurately quantify CHA 
by proposing an approach that allows us to separate charge-asymmetric
and charge-symmetric parts of the hydration free energies of
neutral solutes.
The crux of the approach is the identification
of pairs of neutral solutes that show charge-asymmetric response of
water similar to that that of the K$^+$/F$^-$ pair, based on a set of
quantitative criteria we infer from the behavior of K$^+$/F$^-$ pair
with respect to the charge-symmetry breaking perturbation.  For this
purpose we use a combination of ``CHA-aware'' implicit solvation model
and free energy perturbation simulations in different explicit water
models to search through a large, comprehensive set of small, neutral
drug-like molecules. The search has yielded two ``CHA-conjugate'' pairs
of neutral molecules that behave just like the K$^+$/F$^-$ pair with
respect to asymmetric charge hydration. Unlike the corresponding quantity
for K$^+$/F$^-$ pair, the measured relative hydration asymmetry of
neutral solutes is very accurate. The difference between
experimental hydration energies within each pair of these special 
neutral molecules,
relative to the pair's average hydration energy, quantifies the
intrinsic charge-asymmetric response of real water, free of extrinsic,
complicating factors such as the energetic cost of 
charge species to cross 
the liquid/vapor boundary. 

Quantitatively, we find the asymmetry of the 
water response to hydrated microscopic charge (size about 1 \AA) 
to be close to one-half of the average charge
hydration energy of the charge itself, which means that the charge-asymmetry of the
intrinsic aqueous response to microscopic fields is strong. Given 
the paucity 
of available experimental characteristics of
electrostatic properties of water molecule in liquid phase at
microscopic level, this result is important. 

    The availability of a novel, accurate reference value for the charge
hydration asymmetry made possible through this work, should be of
interest in its own right. For example, we have found that the ability
of an explicit water model to predict the correct experimental relative
CHA for just one pair of neutral solutes correlates well with the
accuracy of the model in predicting absolute hydration energies of small
neutral molecules covering a  wide range of hydration energies.  The
observation suggests an immediate use of the proposed neutral
molecule-based CHA reference for testing, and ultimately improving,
solvent models. Note that computational estimate of hydration energies
of small neutral solutes is now inexpensive, straightforward and  free
from serious complicating issues associated with analogous computations
for charged species. The inclusion of the proposed accurate CHA
reference as an additional optimization target in the process of water
model construction may eventually result in more accurate models of
water. 

\section{Computational Methods}
\subsection{The CHA-GB equation}
The implicit solvent hydration free energies were computed using the CHA
aware generalized Born equation~\cite{Mukhopadhyay2014}

\begin{equation} 
  \Delta G= -\frac{1}{2}\faco\sum_{i,j}\frac{q_iq_j}{f^{CHA-GB}_{ij}},
\label{eq:chagb} 
\end{equation}

\noindent
where $q_i$ is the charge of atom $i$,
$f^{CHA-GB}_{ij}=\sqrt{r_{ij}^2+\tilde{R}_i\tilde{R}_je^{-\frac{r_{ij}^2}{4R_iR_j}}}$
: $R_i$ is its effective Born radius, $r_{ij}$ is the spatial separation
between atom $i$ and $j$, $\epsilon=80$ is the dielectric constant of
water. Here $\tilde R_i=R_i\asymgbinv$ introduces the CHA scaling of the
effective Born radii, $R_i$ analogous to the CHA scaling of the Born
radii in \ref{eq:bornreborn}. $\tau$ is a positive constant
$\mathcal{O}(1)$ that controls the effective range of the neighboring
charges ($j$) affecting the CHA of atom ($i$). The effective Born radii
$R_i$ are computed using the numerical implementation of the ``R6''
formula~\cite{Grycuk03}, over the dielectric surface (boundary).  The
later is obtained by increasing~\cite{Chan1979} the intrinsic atomic
radii by $R_s = 0.52$~\AA~, see Ref~\cite{Mukhopadhyay2014}, and further
rolling a probe of radius $R_w-R_s$, where $R_w$ = 1.4~\AA~
defines the size of water molecule.

\subsection{Selection of small molecules and identification  CHA-conjugate pairs}
To identify  CHA-conjugate molecule pairs from the original 504
molecule set~\cite{Mobley2009} we first selected 250 most rigid
molecules that undergo very little ($<$0.3~\AA) conformational change,
as seen from molecular dynamics trajectories~\cite{Mobley2008b}.  We
then picked 60 ``polar dominant'' molecules -- those with the smallest
nonpolar $\Delta G$ to polar $\Delta G$ ratio ($<$0.2), based on
previous solvation energy estimates~\cite{Mobley2009} in TIP3P water.
We first performed single point $\Delta G$ estimates using a CHA-aware
implicit solvation model; the polar part was modeled using the CHA-GB
equation \ref{eq:chagb}, and the nonpolar part modeled as
$\gamma SASA$, where $SASA$ is the solvent accessible surface area
computed numerically~\cite{MSMS} and $\gamma$ = 0.005 kcal/mol/\AA$^2$.
The topology and coordinates for the molecules were obtained from
Ref.~\cite{Mobley2009}.  Intrinsic atomic radii set and $\tau$,
\ref{eq:chagb}, were optimized (using Nelder-Mead simplex
algorithm~\cite{Nelder1965}) against the experimental solvation free
energy for the 60 polar molecules using different values of $\delta$ in
the range (0, 1.0~\AA). For each value of $\delta$ we performed 2000
separate optimizations using random initial guess; the parameter set
corresponding to the smallest deviation from the experiments, was used for
analysis. The resulting $\Delta G$ for different values of $\delta$ were
used to shortlist an initial selection of 11 molecule pairs (17 distinct
molecules) that showed promising CHA-conjugate behavior with respect to
the  $\Delta \Delta G (\delta)$ gap.  The initial set was then refined
via careful explicit water free energy perturbation calculations in
TIP3P, TIP4P-Ew and TIP5P, resulting in  two CHA-conjugate pairs,
Table~\ref{tab:eta0} that exhibit near perfect symmetric, monotonically
increasing $\Delta \Delta G (\delta)$ gap.

\subsection{Explicit solvation free energies}
Molecule topology and coordinate files were prepared in an earlier
work~\cite{Mobley2009}, using GAFF~\cite{Wang2004} small molecule
parameters assigned by Antechamber 14~\cite{Wang2006}, and the partial
charges were assigned using the Merck-Frosst implementation of
AM1-BCC~\cite{Jakalian2002}. The hydration free energy calculations in
explicit water were performed in GROMACS 4.6.5~\cite{gromacs} using
standard free energy perturbation (FEP) calculations~\cite{Mobley2009}
-- the coulomb and van der Waals coupling was reduced from 1 to 0 using
20 intermediate $\lambda$ values. Molecules were solvated in triclinic
box with at least 12~\AA\ from the solute to the nearest box edge. Real
space electrostatic cutoff was 10~\AA, and long-range electrostatic
interactions were calculated using periodic boundary conditions {\it
via.} the particle mesh Ewald (PME) summation~\cite{essmann95,darden93}
and all bonds were restrained using the LINCS algorithm. Production
simulations were 5 ns in length at each $\lambda$ value, and free
energies and the associated uncertainties were computed using the
Bennett acceptance ratio (BAR), namely the {\it gbar} feature in GROMACS
4.6.5. The FEP estimates were validated by ensuring convergence of
forward (turning the coupling on) and backward (turning the coupling off)
computations for two molecules, and also by comparing with the FEP
estimates for TIP3P in Ref.~\cite{Mobley2009}. The average errors
computed using {\it gbar} were roughly 0.05 kcal/mol (see SI); the
forward and backward computations were
comparable with a difference of 0.04 kcal/mol for the two molecules.  The
same topology and coordinate files were used along with identical
simulation protocol to perform the FEP calculations in each of the four
water models, TIP5P, TIP3P, TIP4P-Ew and OPC~\cite{Anandakrishnan2013}
for 17 molecules (including the 11 shortlisted pairs, above). FEP
calculations were also performed for 3 additional molecules, such that
the set of these 20 molecules span the range of $\Delta G$ from  -0.7 to
-10.0 kcal/mol, which is close to the entire experimental range seen in
the set of 504 small neutral molecules. The estimated $\Delta G$ for the
set of 20 molecules was used to compare the four water models against
the experimental $\eta*$.  Their estimated solvation free energies are
provided in SI.

\begin{acknowledgments}
This work has been supported by NIH grant R01 GM076121
and, in part, by NSF grant CNS-0960081 and the HokieSpeed
supercomputer at Virginia Tech.
\end{acknowledgments}
\providecommand*\mcitethebibliography{\thebibliography}
\csname @ifundefined\endcsname{endmcitethebibliography}
  {\let\endmcitethebibliography\endthebibliography}{}

\end{document}